\PassOptionsToPackage{dvipsnames}{xcolor} 
\documentclass[runningheads]{llncs}
\usepackage[dvipsnames]{xcolor}
\usepackage{amsmath, amssymb}   
\usepackage{caption}
\usepackage{verbatim}
\usepackage{paralist}
\usepackage{xspace}
\usepackage{tikz,ifthen,pgfplots}
\usetikzlibrary{arrows,trees,backgrounds,automata,shapes,decorations,plotmarks,fit,calc,positioning,shadows,chains}
\tikzstyle{every pin edge}=[<-,shorten <=1pt]
\tikzstyle{neuron}=[circle,fill=black!25,minimum size=17pt,inner sep=0pt]
\tikzstyle{input neuron}=[neuron, fill=green!50]
\tikzstyle{output neuron}=[neuron, fill=red!50]
\tikzstyle{hidden neuron}=[neuron, fill=blue!50]
\tikzstyle{annot} = [text width=6em, text centered]
\tikzstyle{nnedge} = [-{stealth},shorten >=0.1cm, shorten <=0.05cm,line width=0.8pt,black]
\tikzstyle{proofNode} = [rounded rectangle, fill=red!30]
\tikzstyle{proofLeaf} = [rounded rectangle, fill=orange!50]
\tikzstyle{proofEdge} = [-{stealth},shorten >=0.1cm, shorten <=0.05cm,line width=0.8pt,black]
\usetikzlibrary{calc}
\newcommand{\relu}{\text{ReLU}\xspace{}}
\newcommand{\sat}{\texttt{SAT}\xspace}
\newcommand{\unsat}{\texttt{UNSAT}\xspace}
\newcommand{\farkasUpper} [1] {f_u(#1)}
\newcommand{\farkasLower} [1] {f_l(#1)}

\usepackage[numbers]{natbib}
\usepackage{listings}
\lstdefinestyle{imandra} {
    aboveskip=5pt,
    language=caml,
    keywordstyle=\color{blue},
    commentstyle=\itshape\color{purple},
    stringstyle=\color{brown},
    basicstyle=\ttfamily\footnotesize,
    breakatwhitespace=false,         
    breaklines=true,                 
    captionpos=b,       
    keepspaces=true,
    showspaces=false,                
    showstringspaces=false,
    showtabs=false,                  
    tabsize=2,
    inputencoding=utf8,
    extendedchars=true,
    literate={φ}{{$\phi$}}1
}

\usepackage{adjustbox}
\newsavebox{\lstbox}

\usepackage{array}
\usepackage{multirow}
\usepackage{booktabs}
\newcolumntype{P}[1]{>{\centering\arraybackslash}p{#1}}

\makeatletter
\renewcommand*{\@fnsymbol}[1]{\ensuremath{\ifcase#1\or *\or \dagger\or \ddagger\or
   \mathsection\or \mathparagraph\or \|\or **\or \dagger\dagger
   \or \ddagger\ddagger \else\@ctrerr\fi}}
\makeatother

\begin{document}

\title{Towards a Certified Proof Checker for Deep Neural Network Verification}

\author{Remi Desmartin\inst{1}\thanks{Both authors contributed equally}\thanks{Funded by Imandra Inc.} \and
    Omri Isac\inst{2}$^*$\and
    Grant Passmore\inst{3} \and
    Kathrin Stark\inst{1}\and
    Guy Katz\inst{2} \and
    Ekaterina Komendantskaya\inst{2}\thanks{Funded by EPSRC grant AISEC (EP/T026952/1) and NCSC grant ``Neural Network Verification: in search of the missing spec.'' }
}


\authorrunning{R. Desmartin, O. Isac  et al.}
%
\institute{ 
Heriot-Watt University, Edinburgh
\and The Hebrew University of Jerusalem
\and Imandra Inc., Austin, Texas, USA}
\maketitle              
\begin{abstract}
Recent developments in deep neural networks (DNNs) have led to their adoption in safety-critical systems, which in turn has heightened the need for guaranteeing their safety. These safety properties of DNNs can be proven using tools developed by the verification community. However, these tools are themselves prone to implementation bugs and numerical stability problems, which make their reliability questionable. To overcome this, some verifiers produce proofs of their results which can be checked by a trusted checker.
In this work, we present a novel implementation of a proof checker for DNN verification. It improves on existing implementations by offering numerical stability and greater verifiability. To achieve this, we leverage two key capabilities of Imandra, an industrial theorem prover: its support of infinite precision real arithmetic and its formal verification infrastructure. So far, we have implemented a proof checker in Imandra, specified its correctness properties and started to verify the checker's compliance with them. Our ongoing work focuses on completing the formal verification of the checker and further optimizing its performance.

\end{abstract}
\section{Introduction}
\label{sec:introduction}

Applications of deep neural networks (DNNs) have grown rapidly in recent years, as they are able to solve computationally hard problems. This has led to their wide use in safety-critical applications like medical imaging \cite{suzukiOverviewDeepLearning2017} or autonomous aircraft \cite{julianDeepNeuralNetwork2019}.
However, DNNs are hard to trust for safety-critical tasks, notably because small perturbations in their inputs -- whether from faulty sensors or malicious adversarial attacks -- may cause large variations of their outputs, leading to potentially catastrophic system failures \cite{szegedyIntriguingPropertiesNeural2013}. To circumvent this issue, the verification community has developed techniques to guarantee DNN correctness using formal verification, employing mathematically rigorous techniques to analyze DNNs' possible behaviours in order to prove it safe and compliant e.g.~\cite{wangBetaCROWNEfficientBound2021, ferrari2022complete, brixDebonaDecoupledBoundary2020, prabhakarAbstractionBasedOutput2019, ferlezFastBATLLNNFast2021, KaHuIbJuLaLiShThWuZeDiKoBa19, bakNnenumVerificationReLU2021, khedrPEREGRiNNPenalizedRelaxationGreedy2021, smithRefutationbasedAdversarialRobustness2021, henriksenDEEPSPLITEfficientSplitting2021}. 
Along with these DNN verifiers, the community created an annual competition~\cite{BrMuBaJoLi23} and a standardisation of an ad-hoc format~\cite{BaKaGuPuNaTa19}.

Usually, DNN verifiers consider an optimized DNN and prove input-output properties, e.g., that for inputs within a delimited region of the input space, the network's output will be in a safe set. 
Besides verifying DNNs at a component level, verification has the power to verify larger systems integrating DNNs. Integration of DNN verifiers in larger verification frameworks has been studied as well~\cite{daggittVehicleHighLevelLanguage}, and it requires the DNN verifiers to provide results that can be checked by the system-level verifier.

Unfortunately, DNN verifiers are susceptible to errors as any other program. One source of problems is floating-point arithmetic used for their internal calculations. While crucial for performance, floating-point arithmetic also leads to numerical instability and is known to compromise soundness~\cite{JiRi21}. As the reliability of DNN verifiers becomes questionable, it is necessary to check that their results are not erroneous. 
When a DNN verifier concludes there exists a counterexample for a given property, this result can be easily checked by evaluating the counterexample over the network and ensuring the property's violation. However, when a verifier concludes that no counterexample exists, ensuring the correctness of this result becomes more complicated. 

To overcome this, DNN verifiers may produce proofs for their results, allowing an external program to check their soundness. Producing proofs is a common practice~\cite{Ne98, BaDeFo15}, and was recently implemented on top of the Marabou DNN verifier~\cite{KaHuIbJuLaLiShThWuZeDiKoBa19, IsBaZhKa22}. 
Typically, proof checkers are simpler programs than the DNN verifiers, and hence much easier to inspect and verify. Moreover, while verifiers are usually implemented in performance-oriented languages such as C++, trusted proof checkers could be implemented in languages suitable for verification.


Functional programming languages (FPL), such as Haskell, OCaml and Lisp are well-suited for this task, thanks
to their deep relationship with logics employed by theorem provers. In fact, some FPLs, such as Agda~\cite{norell2009dependently}, Coq~\cite{CoqProofAssistanta}, ACL2~\cite{kaufmann1996acl2}, Isabelle~\cite{paulson1994isabelle} and Imandra~\cite{passmoreImandraAutomatedReasoning2020} are also theorem provers in their own right.
Implementing and then verifying a program in such a theorem prover allows to bridge the verification gap, i.e. minimise the discrepancies that can exist between the original
(executable) program and its verified (abstract) model~\cite{breitnerReadySetVerify2021}.

In this paper, we describe our ongoing work to design, implement and verify a \emph{formally-verifiable and infinitely-precise proof checker} for DNN verifiers. We have implemented an adaptation of a checker of \unsat proofs produced by the Marabou DNN verifier~\cite{KaHuIbJuLaLiShThWuZeDiKoBa19,IsBaZhKa22} to Imandra~\cite{passmoreImandraAutomatedReasoning2020}, a functional programming language coupled with its own industrial-strength theorem prover.
Three key features make Imandra a suitable tool: infinite precision real arithmetic, efficient code extraction and the first-class integration of formal verification. Support for infinite precision real arithmetic prevents numerical instability. The ability to extract verified Imandra code to native OCaml improves scalability as it can then benefit from the standard OCaml compiler's optimizations. Finally, with Imandra's integrated formal verification, we can directly analyze the correctness of the proof checker we implement. Note that Imandra as a DNN verifier has already been researched~\cite{desmartinNeuralNetworksImandra2022}.
\vspace{-0.4cm}
\paragraph{Contributions.} Contrary to previous implementations prioritising scalability, our checker can be formally verified by Imandra's prover and its precision is infinite. This increases the checker's reliability and overcomes a main barrier in integrating DNN verifiers in system-level checkers. Since reliability usually compromises scalability, our
proof checker supports several checking modes, with different approaches to balance the two. This is done along two orthogonal axes, by optionally:
\begin{inparaenum}[(i)]
    \item using verified data structures at the expense of computation speed;
    \item accepting some parts of the proof without checking. 
\end{inparaenum}

Our ongoing work is currently focused on formally verifying the proof checker. So far, we have managed to verify that our checker complies with linear algebra theorems, and we attempt to leverage these results to verify the proof checker as a whole in the future.

\vspace{-0.4cm}
\paragraph{Paper organisation.} The rest of this paper is organized as follows. In Section~\ref{sec:background} we provide relevant background on DNN verification and proof production. In Section~\ref{sec:construction} and Section~\ref{sec:verification} we respectively describe our proof checker, and our ongoing work towards formally verifying it using Imandra. In Section~\ref{sec:conclusion} we conclude our work, and describe our plans for completing our work and for the future.



\section{Background}
\label{sec:background}
\subsection{DNN Verification}
\label{sec:DNNverification}
\begin{figure}[h]
  \begin{center}
  \def\layersep{1.5cm}
  \scalebox{0.85}{
      \begin{tikzpicture}[shorten >=1pt,->,draw=black!50, node distance=\layersep,font=\footnotesize]
        
        \node[input neuron] (I-1) at (0,-1) {$x_1$};
        \node[input neuron] (I-2) at (0,-2) {$x_2$};
    
        \node[hidden neuron] (H-1) at (\layersep, -1) {$v_1$};
        \node[hidden neuron] (H-2) at (\layersep,-2) {$v_2$};
        \node[hidden neuron] (H-3) at (2*\layersep,-1.5) {$v_3$};
        \node[output neuron] (O-1) at (3*\layersep,-1.5) {$y$};
        
        \draw[nnedge] (I-1) --node[above,pos=0.4] {$2$} (H-1);
        \draw[nnedge] (I-2) --node[above,pos=0.4] {$1$} (H-2);
        
        \draw[nnedge] (H-1) --node[above] {$-1$} (H-3);
        \draw[nnedge] (H-2) --node[below] {$1$} (H-3);
    
        \draw[nnedge] (H-3) --node[above] {$1$} (O-1);
        
        \node[above=0.01cm of H-1] (b1) {$\relu$};
        \node[above=0.01cm of H-2] (b1) {$\relu$};
        \node[above=0.01cm of H-3] (b1) {$\relu$};
      \end{tikzpicture}
  }
\end{center}
\vspace{-0.2cm}

\caption{A Simple DNN. The bias parameters are all set to zero and are ignored. }
\vspace{-0.5cm}
\label{fig:dnnexample}
\end{figure}
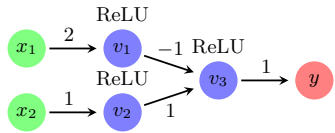

Throughout the paper, we focus on DNNs with ReLU activation functions, though all our work can be extended to DNNs using any piecewise-linear activation functions (e.g.,
\emph{max pooling}). We refer the reader to Appendix~\ref{app:dnn} for a formal definition of DNNs and activation functions.
An example of a DNN appears in Fig.~\ref{fig:dnnexample}.

\vspace{-0.5cm}
The \emph{DNN verification problem} is the decision problem of deciding whether for a given DNN 
$\mathcal{N}:\mathbb{R}^m\rightarrow\mathbb{R}^k $ and a property $P\subseteq\mathbb{R}^{m+k}$, there exists an input $x\in\mathbb{R}^m$ such that $\mathcal{N}(x)=y \wedge P(x,y)$. If such $x$ exists, the verification query is
\textit{satisfiable} (\sat); otherwise it is
\textit{unsatisfiable} (\unsat). Typically, $ P $ represents an erroneous behaviour, thus
an input $x$ satisfying the query serves as a counterexample and 
\unsat{} indicates the network acts as expected. 

Due to its linear and piecewise-linear structure, a DNN verification query can be reduced to an instance of  Linear Programming (LP)~\cite{Da63}, representing the affine functions of the DNN, and 
piecewise-linear constraints that represent the activation functions and the property.  This reduction makes algorithms for solving LP instances, coupled with a \emph{case-splitting} approach for handling
the piecewise-linear constraints~\cite{BaIoLaVyNoCr16, KaBaDiJuKo21}, a prime scheme for DNN verification, which we call \emph{LP-based DNN verifiers}. 

The widely used Simplex algorithm~\cite{Da63, KaBaDiJuKo21,DuDe06}, is typically used by such verifiers.
Based on the problem constraints, the algorithm initiates a matrix $A$ called the \emph{tableau}, a variable vector $x$ and two \emph{bound vectors} $u,l$ such that $l\leq x \leq u$. The Simplex algorithm then attempts to find a solution to the system:
\begin{equation}
Ax = 0 \wedge l \leq x \leq u    
\end{equation}
or concludes that none exists. For clarity, we denote $u(x_i),l(x_i)$ as the upper and lower bounds of the variable $x_i$, instead of $u_i,l_i$.
\vspace{-0.2cm}
\begin{example}
\label{ex1}
Consider the DNN in Fig.~\ref{fig:dnnexample} and the property $P$ that holds if and only if $(x_1, x_2)\in [-1,1]^2 \wedge y\in[2,3]$. We later show a proof of \unsat{} for this query.
We assign variables $x_1,\:x_2,\:y$ to the input and output neurons. For all $i\in {1,2,3}$ we assign a couple of variables $f_i,b_i$ for the inputs and outputs of the neurons $v_i$, where $f_i=\relu(b_i)$. We then get the linear constraints and bounds (where some bounds were arbitrarily fixed for simplicity): 
\vspace{-0.2cm}
\begin{equation}
b_1 = 2x_1,\:\: b_2 = x_2,\:\: b_3 = f_2 - f_1,\:\: y = f_3    
\end{equation}
\begin{equation}
    -1\leq x_1, x_2, b_2 \leq 1,\: 0 \leq f_2 \leq 1,\: -2\leq b_1,b_3 \leq 2, \: 0 \leq f_1, f_3 \leq 2, \: 2 \leq y \leq 3
\end{equation} 
and the piecewise linear constraints: $\forall i \in {1,2,3}: f_i = \relu(b_i) $

Then, an LP-based DNN verifier initiates the input for the Simplex algorithm:

\begin{tabular}{ c c } 
 $A =
  \begin{bmatrix}
    2 \:\:  & \: 0 \:  & -1  & 0  & 0  & 0  & \: 0 \: & 0 &\:\: 0 \\
    0 \:\:  & \: 1 \:  &  0 & -1 & 0  & 0  & \: 0 \: & 0 & \:\:0 \\
    0 \:\:  & \: 0 \:  &  0 &  0 & -1 & -1 & \: 1 \: & 0 &\:\: 0 \\
    0 \:\:  &  \:0 \:  &  0 & 0  & 0  & 0  &\:  0 \: & -1 &\:\: 1 \\
    \end{bmatrix} $

    \begin{tabular} {c}
         $u =
            \begin{bmatrix}
                \:\:\:1 & \:\:\: 1 &\:\:\:\: 2 &\:\:\:\: 1 & \:\:\:\: 2 &\:\:\:\: 2 &\:\:\:\: 1 &\:\:\:\: 3 &\:\:\:\: 3
            \end{bmatrix}^\intercal $ \\[0.1cm]
        $x =
            \begin{bmatrix}
                \:\:x_1 &\:\: x_2 &\:\: b_1 &\:\: b_2 &\:\:b_3 &\:\:f_1&\:\: f_2 & \:\:f_3&\:\: y 
            \end{bmatrix}^\intercal$ \\[0.1cm]
        $l = \:
            \begin{bmatrix}
                 -1 & -1 & -2 & -1 & -2 &\:\:\:\: 0 &\:\:\:\: 0 &\:\:\:\: 0 &\:\:\:\: 2
            \end{bmatrix} ^\intercal$ 
           
    \end{tabular}
\end{tabular}
        
In addition to the piecewise-linear constraints $ \forall i \in {1,2,3}: f_i = \relu(b_i)$.
\end{example}


%
\vspace{-0.2cm}

One of the key tools used by the Simplex algorithm, and consequently by DNN verifiers, is dynamic bound tightening. This procedure allows deducing tighter bounds for each variable and is crucial for the solver's performance. For example, using the above equation $f_3=y$ and the bound $u(y) = 2$, we can deduce $u(f_3)= 2$, and further use this bound to deduce other bounds as well. 
The piecewise-linear constraints introduce rules for tightening bounds as well, which we call \emph{Theory-lemmas}.
For instance, the output variable $f_3$ of the \relu{} constraint of the above example is upper bounded by the input variable $b_3$, whose upper bound is 2. The list of supported lemmas appears in Appendix~\ref{app:lemmachecking}. 

The case-splitting approach is used over the linear pieces of some piecewise-linear constraints, creating several sub-queries with each adding new information to the Simplex algorithm.  For example, when performing a split over a constraint of the form $y=\relu(x)$, two sub-queries are created. One is enhanced with  $y=x\wedge x\geq0$, and the other with $y=0\wedge x\leq0$.
The use of case-splitting also induces a tree structure for the verification algorithm, with nodes corresponding to the splits
applied. On every node, the verifier attempts to conclude the satisfiability of the query based on its linear constraints. If it concludes an answer, then this node represents a leaf.
In particular, a tree with all leaves corresponding to an \unsat{} result of Simplex is a search tree of an \unsat{} verification query.
\vspace{-0.2cm}
\subsection{Proof Production for DNN Verification}
\label{sec:proofs}
Proof production for \sat{} is straightforward using a satisfying assignment. On the other hand, when a query is \unsat{}, the verification algorithm induces a search tree, where each leaf corresponds to
an \unsat{} result of the Simplex algorithm for that particular leaf. Thus, a proof of \unsat{} is comprised of a matching proof tree where each leaf contains a proof of the matching  Simplex  \unsat{} result. 
Proving \unsat{} results of Simplex is based on a constructive version of the Farkas Lemma~\cite{Va96}, which identifies the proof for \unsat{} LP instances.
Formally, it was proven~\cite{IsBaZhKa22} that:
\begin{theorem}
\label{thm:mainthm}

    	Let $ A \in M_{m \times n} (\mathbb{R})$  and $ l,x,u \in \mathbb{R}^n $, such that $ A\cdot x = 0 $ and $ l \leq x \leq u$,	
	exactly one of these two options holds:
	\begin{enumerate}
		\item The \sat{} case:
		$ \exists x \in \mathbb{R}^n $ such that $ A \cdot x =
		0 $ and $ l \leq x \leq u $. 
		\item The \unsat{} case: $ \exists w \in \mathbb{R}^m $ such
		that for all  $ l \leq x \leq u $,  $ w^\intercal
                \cdot A \cdot x < 0$, whereas $ 0 \cdot w = 0 $. Thus,
                $w$ is a proof of the constraints' unsatisfiability.
              \end{enumerate}
 	Moreover, these vectors can be constructed while executing the Simplex algorithm.
\end{theorem}
\vspace{-0.2cm}

To construct the proof vectors, two column vectors are assigned to each variable $x_i$, denoted $\farkasUpper{x_i}, \farkasLower{x_i}$, which are updated during bound tightening. These vectors are used to prove the tightest upper and lower bounds of $x_i$ deduced during the bound tightenings performed by Simplex, based on $u,l$ and $A$. This mechanism was designed and implemented~\cite{IsBaZhKa22}, on top of the Marabou DNN verifier~\cite{KaHuIbJuLaLiShThWuZeDiKoBa19}.

Supporting the complete tree structure of the verification algorithm is done by constructing the proof tree in a similar manner to the search tree --- every split performed in the search directly creates a similar split in the proof tree, with updates to the equations and bounds introduced by the split.
Proving theory lemmas is done by keeping details about the bound that invoked the lemma together with a Farkas vector proving its deduction and the new learned bound, and adding them to the corresponding proof tree node.

\section{The Imandra Proof Checker}
\label{sec:construction}
Our proof checker is designed to check proofs produced by the Marabou DNN verifier~\cite{KaHuIbJuLaLiShThWuZeDiKoBa19}, to the best of our knowledge the only proof producing DNN verifier.
When given a Marabou proof of \unsat{} as a JSON~\cite{Br14} file, the proof checker reconstructs the proof tree using datatypes encoded in Imandra.

The proof tree consists of two different node types --- a proof node and a proof leaf. Both node types contain a list of lemmas and a corresponding split. In addition, a node contains a list of its children, and a leaf contains a contradiction vector, as constructed by Theorem~\ref{thm:mainthm}. This enables the checker to check the proof tree structure at the type-level.
The proof checker also initiates a matrix $A$ called a \emph{tableau}, vectors of upper and lower bounds $u,l$ and a list of piecewise-linear constraints (see Section~\ref{sec:DNNverification}).

The checking process consists of traversing the proof tree. For each node, the checker begins by locally updating $u,l$  and $A$ according to the split, and optionally checking the correctness of all lemmas. Lemma checking is similar to checking contradictions, as shown in Example~\ref{ex:proof_checking} below (see Appendix~\ref{app:lemmachecking} for details).

If the node checked is not a leaf, then the checker will check that all its childrens' splits correspond to some piecewise-linear constraint of the problem i.e. one child has a split of the form $y=x\wedge x\geq0$ and the other of the form $y=0\wedge x\leq0$ for some constraint $y=\relu(x)$. If the checker certifies the node, it will recursively check all its children, passing changes to $u,l$ and $A$ to them.

When checking a leaf, the checker checks that the contradiction vector $w$ implies \unsat{}, as stated in Theorem~\ref{thm:mainthm}. As implied from the theorem, the checker will first create the row vector $w^\intercal \cdot A$, and will compute the upper bound of its underlying linear combination of variables $w^\intercal \cdot A \cdot x$. The checker concludes by asserting this upper bound is negative.

The checker then concludes that the proof tree represents a correct proof if and only if all nodes passed the checking process.
\begin{example}\label{ex:proof_checking}
Consider the simple proof in Fig.~\ref{fig:example}. The root contains a single lemma and each leaf contains a contradiction vector, which means the verifier performed a single split. In addition, the proof object contains the tableau $A$, the bound vectors $u,l$, and the \relu{} constraints as presented in Example~\ref{ex1}.
\vspace{-0.5cm}
\begin{figure}[ht]
	\centering
	\scalebox{0.85} {
		\def\xSep{2cm}
		\def\ySep{1.2cm}
		\begin{tikzpicture}[ >=stealth,shorten >=1pt,shorten <=1pt]
			
			\node[proofNode] (root) at (0,0) []  {$A,l,u$};

            \node[below=0.01cm of root] (lemma) [] {$u(b_3)=1\rightarrow u(f_3)=1$,
                $ \begin{bmatrix}
                    0&0&1&0
            \end{bmatrix}^\intercal$};
   
			\node[proofLeaf] (active) at (-\xSep, -1.5*\ySep)  [] {$(f_3=b_3)\wedge (b_3\geq 0)$};
			\node[proofLeaf] (inactive) at (\xSep, -1.5*\ySep)  []  {$(f_3=0)\wedge (b_3\leq 0)$};

			\draw[proofEdge] (lemma) --node[label={[xshift=-0.4cm,yshift=-0.1cm]}] {} (active);
			\draw[proofEdge] (lemma) --node[label={[xshift=0.4cm,yshift=-0.1cm]}] {} (inactive);
		
			\node[below=0.05cm of inactive] {$\begin{bmatrix} 0&0&0&1\end{bmatrix}^\intercal$};
   
			\node[below=0.05cm of active] {$\begin{bmatrix} 0&0&1&-1&1\end{bmatrix}^\intercal$};

		\end{tikzpicture}
	}
     \vspace{-0.2cm}
	\caption{A proof tree example.}
	\label{fig:example}
\end{figure}
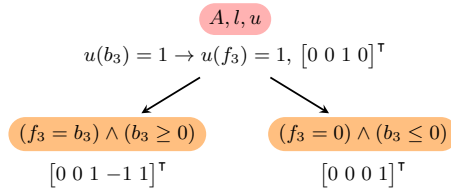

The proof checker begins by checking the lemma of the root. It does so by creating the linear combination $\begin{bmatrix}
    0&0&1&0
\end{bmatrix}^\intercal\cdot A \cdot x = -b_3-f_1+f_2$. As the lemma is invoked by the upper bound of $b_3$, the checker uses the equivalent equation $b_3 = f_2 - f_1$, which gives the upper bound $u(b_3) = u(f_2)-l(f_1) = 1$. We can indeed deduce the bound $u(f_3) = 1$ based on the constraint $f_3=\relu(b_3)$, so the lemma proof is correct.
Then, the checker certifies that the splits $f_3 = 0 \wedge b_3 \leq 0$ and $f_3 = b_3 \wedge b_3 \geq 0$ correspond to the two splits of $f_3=\relu(b_3)$.
The checker then begins checking the left leaf. It starts by updating $l(b_3)=0$ and adding the equation $f_3 = b_3$ as the row $\begin{bmatrix}0&0&0&0&1&0&0&-1&0\end{bmatrix}$ to $A$. Then, the checker checks the contradiction vector by computing $\begin{bmatrix}
    0&0&1&-1&1
\end{bmatrix}^\intercal\cdot A \cdot x = -f_1+f_2-y$.  The upper bound of this combination is $-l(f_1)+u(f_2)-l(y)=-1$ which is negative, thus proving \unsat{} for the leaf according to Theorem~\ref{thm:mainthm}.
Checking the right leaf is done similarly. After checking all nodes, the checker asserts the proof tree indeed proves \unsat{} for the whole query.
\vspace{-0.3cm}
\end{example}

\subsubsection{Implementation in Imandra, OCaml Extraction and Evaluation.}\label{subsec:impl}
Porting the proof checker from C++ to Imandra necessitates taking into account the trade-off between scalability and computation.

The choice of data structures for common objects -- like vectors -- is essential in the balance between scalability and efficiency~\cite{desmartinNeuralNetworksImandra2022}. In this work, we experiment with two different implementations for vectors: native OCaml lists, and sparse vectors using Imandra's built-in \lstinline{Map} data type, based on binary search trees. The latter has better performance but the former makes it easier to verify, so for now our verification efforts focus on the native list implementation (see Appendix~\ref{app:imandra} for more details). 

Imandra's logic includes theories for arbitrary precision integer and real arithmetic, which are implemented using OCaml's built-in \lstinline{Zarith} library~\cite{mineZarithLibrary2023}. As a result, the Imandra implementation of the checker supports arbitrary precision real arithmetic with low overhead. 

Executing code within Imandra's reasoning environment is helpful during the implementation and verification process, but is not optimized for performance. To that end, \lstinline{imandra-extract} is a facility to extract native OCaml code that can be compiled -- and optimized -- with standard OCaml compilers. The extracted code retains Imandra's semantics, meaning that it still uses infinite precision real arithmetic.
An initial comparison of the execution time for checking the same proofs from the ACAS-Xu benchmark~\cite{KaHuIbJuLaLiShThWuZeDiKoBa19} in the C++ implementation and in the extracted OCaml code with native lists shows that our implementation is about 150 times slower than the original implementation but stays within a reasonable time, i.e. less than 40 minutes for all the examples ran (see Appendix~\ref{app:evaluation}). Further optimizations and a comprehensive benchmark are ongoing work. 

\section{Specification of the Proof Checker's Correctness}
\label{sec:verification}
We aim to verify the two main checks performed by the proof checker when traversing the proof tree (see Section~\ref{sec:construction}): contradictions and theory lemmas.

\paragraph{Contradictions checking} We want to verify that our proof checker identifies correctly when a contradiction vector is a valid proof of \unsat{}, thus satisfying Theorem~\ref{thm:mainthm} (case 2.). Formally, the specification can be given as:

For all contradiction vector $w$, tableau $A$, bounds $u,l$, and bounded input $l \le x \le u$, if the upper bound of $w^T \cdot A \cdot x$ is negative, then $x$ cannot satisfy the constraints $A \cdot x = 0 \wedge l \leq x\leq u $. The Imandra implementation of this specification is given in Listing \ref{lst:contradiction_thm}.

\begin{lrbox}{\lstbox}
\begin{lstlisting}[
                   frame=single, 
                   caption=\footnotesize{\emph{High-level theorem formalising correctness of contradiction checking. The function \lstinline{check_contradiction} is a key component of the proof checker which should return \lstinline{true} iff the linear combination of the tableau and contradiction vectors has a negative upper bound.}}, 
                   label={lst:contradiction_thm}]
theorem contra_correct x contra tableau u_bounds l_bounds =
  is_bounded x u_bounds l_bounds
  && check_contradiction contra tableau u_bounds l_bounds
   ==> not (null_product tableau x)
\end{lstlisting}
\end{lrbox}
\scalebox{0.9}{\usebox{\lstbox}}
\vspace{-0.2cm}

\paragraph{Theory lemmas} The goal is to prove that each theory lemma within the proof, indeed corresponds to one of the theory lemmas (Appendix~\ref{app:lemmachecking}). 

Proving the specification necessitates guiding Imandra by providing supporting lemmas, in our case properties of linear algebra. After proving these intermediary lemmas, Imandra's proof automation can apply them automatically, or we can manually specify which lemma to apply.

So far we have defined and proved that our checker is coherent with known properties of linear algebra (e.g.  Listing~\ref{lst:verification}). Our current work focuses on building on top of these lemmas to fully prove the checker's correctness.
\vspace{0.2cm}

\begin{lrbox}{\lstbox}
\begin{lstlisting}[frame=single, caption=\footnotesize{\emph{Definition of lemmas proved in Imandra; the first lemma \lstinline{dot_product_coeff} defines the homogeneity of the dot-product operation; it is used to prove the second lemma by using the \lstinline{apply} annotation}}, label={lst:verification}]
lemma dot_product_coeff x y c =
  dot_product x (list_mult y c) = c *. dot_product x y
[@@auto]

lemma dot_product_coeff_eq x y c =
  dot_product x y = 0. ==> dot_product x (list_mult y c) = 0.
[@@auto][@@apply dot_product_coeff x y c]
\end{lstlisting}
\end{lrbox}
\scalebox{0.9}{\usebox{\lstbox}}

\section{Conclusion and Future Work}
\label{sec:conclusion}
We have implemented a checking algorithm for proofs generated by a DNN verifier in the functional programming language of Imandra, enabling the checking algorithm to be infinitely precise and formally verifiable by Imandra's prover. 

Compared to previous work, our implementation presents two new guarantees: it avoids numerical instability by using arbitrary-precision real numbers instead of floating-point numbers; and its correctness can be formally verified as it is implemented in a theorem prover. 
As expected, adding safety guarantees comes at a cost of performance, but the extraction of native OCaml minimises the overhead compared to the unverified C++ implementation.
Furthermore, using an FPL checker to check proofs produced by a DNN verifier is a first step towards integrating component-level DNN verification into the system-level.

Our immediate future work is to continue the verification of the proof checker. In addition, we intend to identify cases where the existing checker implementation fails (e.g. due to numerical instability) and ours correctly checks the proof. Investigating further optimizations is also a promising direction by implementing better performance data structures, such as AVL trees.



\bibliography{main}

\begin{thebibliography}{34}
\providecommand{\natexlab}[1]{#1}
\providecommand{\url}[1]{\texttt{#1}}
\providecommand{\urlprefix}{}

\bibitem[{Coq(1984)}]{CoqProofAssistanta}
{The Coq Proof Assistant} (1984), \url{https://coq.inria.fr}

\bibitem[{Bak(2021)}]{bakNnenumVerificationReLU2021}
Bak, S.: {Nnenum: Verification of ReLU Neural Networks with Optimized
  Abstraction Refinement}.
\newblock In: Proc. 13th Int. Symposioum NASA Formal Methods (NFM). pp. 19--36
  (2021)

\bibitem[{Barrett et~al.(2019)Barrett, Katz, Guidotti, Pulina, Narodytska, and
  Tacchella}]{BaKaGuPuNaTa19}
Barrett, C., Katz, G., Guidotti, D., Pulina, L., Narodytska, N., Tacchella, A.:
  {The Verification of Neural Networks Library (VNN-LIB)} (2019),
  \url{www.vnnlib.org}

\bibitem[{Barrett et~al.(2015)Barrett, de~Moura, and Fontaine}]{BaDeFo15}
Barrett, C., de~Moura, L., Fontaine, P.: {Proofs in Satisfiability Modulo
  Theories}.
\newblock All about Proofs, Proofs for All 55(1), 23--44 (2015)

\bibitem[{Bastani et~al.(2016)Bastani, Ioannou, Lampropoulos, Vytiniotis, Nori,
  and Criminisi}]{BaIoLaVyNoCr16}
Bastani, O., Ioannou, Y., Lampropoulos, L., Vytiniotis, D., Nori, A.,
  Criminisi, A.: {Measuring Neural Net Robustness with Constraints}.
\newblock In: Proc. 30th Conf. on Neural Information Processing Systems
  (NeurIPS) (2016)

\bibitem[{Bray(2014)}]{Br14}
Bray, T.: {The JavaScript Object Notation (JSON) Data Interchange Format}
  (2014), \url{https://www.rfc-editor.org/info/rfc7159}

\bibitem[{Breitner et~al.(2021)Breitner, {Spector-Zabusky}, Li, Rizkallah,
  Wiegley, Cohen, and Weirich}]{breitnerReadySetVerify2021}
Breitner, J., {Spector-Zabusky}, A., Li, Y., Rizkallah, C., Wiegley, J., Cohen,
  J.M., Weirich, S.: {Ready, Set, Verify! Applying Hs-to-Coq to Real-World
  Haskell Code}.
\newblock Journal of Functional Programming 31, e5 (2021)

\bibitem[{Brix et~al.(2023)Brix, M{\"u}ller, Bak, Johnson, and
  Liu}]{BrMuBaJoLi23}
Brix, C., M{\"u}ller, M.N., Bak, S., Johnson, T.T., Liu, C.: {First Three Years
  of the International Verification of Neural Networks Competition (VNN-COMP)}
  (2023), technical report. \url{http://arxiv.org/abs/2301.05815}

\bibitem[{Brix and Noll(2020)}]{brixDebonaDecoupledBoundary2020}
Brix, C., Noll, T.: {Debona: Decoupled Boundary Network Analysis for Tighter
  Bounds and Faster Adversarial Robustness Proofs} (2020), technical report.
  \url{http://arxiv.org/abs/2006.09040}

\bibitem[{Daggitt et~al.(2022)Daggitt, Kokke, Atkey, Arnaboldi, and
  Komendantskaya}]{daggittVehicleHighLevelLanguage}
Daggitt, M.L., Kokke, W., Atkey, R., Arnaboldi, L., Komendantskaya, E.:
  {Vehicle: Interfacing Neural Network Verifiers with Interactive Theorem
  Provers} (2022), technical report. \url{http://arxiv.org/abs/2202.05207}

\bibitem[{Dantzig(1963)}]{Da63}
Dantzig, G.: {Linear Programming and Extensions}.
\newblock Princeton University Press (1963)

\bibitem[{Desmartin et~al.(2022)Desmartin, Passmore, and
  Komendantskaya}]{desmartinNeuralNetworksImandra2022}
Desmartin, R., Passmore, G.O., Komendantskaya, E.: {Neural Networks in Imandra:
  Matrix Representation as a Verification Choice}.
\newblock In: Proc. 5th Int. Workshop of Software Verification and Formal
  Methods for ML-Enabled Autonomous Systems (FoMLAS) and 15th Int. Workshop on
  Numerical Software Verification (NSV). pp. 78--95 (2022)

\bibitem[{Dutertre and de~Moura(2006)}]{DuDe06}
Dutertre, B., de~Moura, L.: {A Fast Linear-Arithmetic Solver for DPLL(T)}.
\newblock In: Proc. 18th Int. Conf. on Computer Aided Verification (CAV). pp.
  81--94 (2006)

\bibitem[{Ferlez et~al.(2022)Ferlez, Khedr, and
  Shoukry}]{ferlezFastBATLLNNFast2021}
Ferlez, J., Khedr, H., Shoukry, Y.: {Fast BATLLNN: Fast Box Analysis of
  Two-Level Lattice Neural Networks}.
\newblock In: Proc. 25th ACM Int. Conference on Hybrid Systems: Computation and
  Control (HSCC). pp. 1--11 (2022)

\bibitem[{Ferrari et~al.(2022)Ferrari, Mueller, Jovanovi{\'c}, and
  Vechev}]{ferrari2022complete}
Ferrari, C., Mueller, M.N., Jovanovi{\'c}, N., Vechev, M.: Complete
  verification via multi-neuron relaxation guided branch-and-bound.
\newblock In: Proc. 10th Int. Conf. on Learning Representations (ICLR) (2022)

\bibitem[{Henriksen and
  Lomuscio(2021)}]{henriksenDEEPSPLITEfficientSplitting2021}
Henriksen, P., Lomuscio, A.: {DEEPSPLIT: An Efficient Splitting Method for
  Neural Network Verification via Indirect Effect Analysis}.
\newblock In: Proc. 30th Int. Joint Conf. on Artificial Intelligence (IJCAI).
  pp. 2549--2555 (2021)

\bibitem[{Isac et~al.(2022)Isac, Barrett, Zhang, and Katz}]{IsBaZhKa22}
Isac, O., Barrett, C., Zhang, M., Katz, G.: {Neural Network Verification with
  Proof Production}.
\newblock In: Proc. 22nd Int. Conf. on Formal Methods in Computer-Aided Design
  (FMCAD). pp. 38--48 (2022)

\bibitem[{Jia and Rinard(2021)}]{JiRi21}
Jia, K., Rinard, M.: {Exploiting Verified Neural Networks via Floating Point
  Numerical Error}.
\newblock In: Proc. 28th Int. Static Analysis Symposium (SAS). pp. 191--205
  (2021)

\bibitem[{Julian et~al.(2019)Julian, Kochenderfer, and
  Owen}]{julianDeepNeuralNetwork2019}
Julian, K., Kochenderfer, M., Owen, M.: {Deep Neural Network Compression for
  Aircraft Collision Avoidance Systems}.
\newblock Journal of Guidance, Control, and Dynamics 42(3), 598--608 (2019)

\bibitem[{Katz et~al.(2021)Katz, Barrett, Dill, Julian, and
  Kochenderfer}]{KaBaDiJuKo21}
Katz, G., Barrett, C., Dill, D., Julian, K., Kochenderfer, M.: {Reluplex: a
  Calculus for Reasoning about Deep Neural Networks}.
\newblock Formal Methods in System Design (FMSD)  (2021)

\bibitem[{Katz et~al.(2019)Katz, Huang, Ibeling, Julian, Lazarus, Lim, Shah,
  Thakoor, Wu, Zelji\'c, Dill, Kochenderfer, and
  Barrett}]{KaHuIbJuLaLiShThWuZeDiKoBa19}
Katz, G., Huang, D., Ibeling, D., Julian, K., Lazarus, C., Lim, R., Shah, P.,
  Thakoor, S., Wu, H., Zelji\'c, A., Dill, D., Kochenderfer, M., Barrett, C.:
  {The Marabou Framework for Verification and Analysis of Deep Neural
  Networks}.
\newblock In: Proc. 31st Int. Conf. on Computer Aided Verification (CAV). pp.
  443--452 (2019)

\bibitem[{Kaufmann and Moore(1996)}]{kaufmann1996acl2}
Kaufmann, M., Moore, J.S.: {ACL2: An Iindustrial Strength Version of Nqthm}.
\newblock In: Proc. 11th Conf. on Computer Assurance (COMPASS). pp. 23--34
  (1996)

\bibitem[{Khedr et~al.(2021)Khedr, Ferlez, and
  Shoukry}]{khedrPEREGRiNNPenalizedRelaxationGreedy2021}
Khedr, H., Ferlez, J., Shoukry, Y.: {PEREGRiNN: Penalized-Relaxation Greedy
  Neural Network Verifier}.
\newblock In: Proc. 33rd Int. Conf. Computer Aided Verification (CAV). pp.
  287--300 (2021)

\bibitem[{Min{\'e} et~al.(2023)Min{\'e}, Leroy, Cuoq, and
  Troestler}]{mineZarithLibrary2023}
Min{\'e}, A., Leroy, X., Cuoq, P., Troestler, C.: {The Zarith Library} (2023),
  \url{https://github.com/ocaml/Zarith}

\bibitem[{Necula(1998)}]{Ne98}
Necula, G.: {Compiling with Proofs}.
\newblock Carnegie Mellon University (1998)

\bibitem[{Norell(2009)}]{norell2009dependently}
Norell, U.: {Dependently typed programming in Agda}.
\newblock In: Proc. 4th Int. workshop on Types in Language Design and
  Implementation (TLDI). pp. 1--2 (2009)

\bibitem[{Passmore et~al.(2020)Passmore, Cruanes, Ignatovich, Aitken, Bray,
  Kagan, Kanishev, Maclean, and
  Mometto}]{passmoreImandraAutomatedReasoning2020}
Passmore, G., Cruanes, S., Ignatovich, D., Aitken, D., Bray, M., Kagan, E.,
  Kanishev, K., Maclean, E., Mometto, N.: {The Imandra Automated Reasoning
  System (System Description)}.
\newblock In: Proc. 10th Int. Joint Conf. Automated Reasoning (IJCAR). pp.
  464--471 (2020)

\bibitem[{Paulson(1994)}]{paulson1994isabelle}
Paulson, L.C.: {Isabelle: A Generic Theorem Prover}.
\newblock Springer (1994)

\bibitem[{Prabhakar and Afzal(2019)}]{prabhakarAbstractionBasedOutput2019}
Prabhakar, P., Afzal, Z.R.: {Abstraction Based Output Range Analysis for Neural
  Networks}.
\newblock In: Proc. 32nd Int. Conf. on Neural Information Processing Systems
  (NeurIPS). pp. 15762--15772 (2019)

\bibitem[{Smith et~al.(2021)Smith, Allen, Swaminathan, and
  Zhang}]{smithRefutationbasedAdversarialRobustness2021}
Smith, J., Allen, J., Swaminathan, V., Zhang, Z.: {Refutation-Based Adversarial
  Robustness Verification of Deep Neural Networks} (2021)

\bibitem[{Suzuki(2017)}]{suzukiOverviewDeepLearning2017}
Suzuki, K.: {Overview of Deep Learning in Medical Imaging}.
\newblock Radiological Physics and Technology 10(3), 257--273 (2017)

\bibitem[{Szegedy et~al.(2013)Szegedy, Zaremba, Sutskever, Bruna, Erhan,
  Goodfellow, and Fergus}]{szegedyIntriguingPropertiesNeural2013}
Szegedy, C., Zaremba, W., Sutskever, I., Bruna, J., Erhan, D., Goodfellow, I.,
  Fergus, R.: {Intriguing Properties of Neural Networks} (2013), technical
  report. \url{http://arxiv.org/abs/1312.6199}

\bibitem[{Vanderbei(1996)}]{Va96}
Vanderbei, R.: {Linear Programming: Foundations and Extensions}.
\newblock Journal of the Operational Research Society  (1996)

\bibitem[{Wang et~al.(2021)Wang, Zhang, Xu, Lin, Jana, Hsieh, and
  Kolter}]{wangBetaCROWNEfficientBound2021}
Wang, S., Zhang, H., Xu, K., Lin, X., Jana, S., Hsieh, C.J., Kolter, J.Z.:
  {Beta-CROWN: Efficient Bound Propagation with Per-neuron Split Constraints
  for Neural Network Robustness Verification}.
\newblock Advances in Neural Information Processing Systems 34, 29909--29921
  (2021)

\end{thebibliography}

\newpage

\newpage

{\noindent\huge{Appendix}}

\renewcommand\thesection{\Alph{section}}
\renewcommand\thesubsection{\thesection.\arabic{subsection}}
\setcounter{section}{0}
\section{Deep Neural Networks}
\label{app:dnn}

Formally, a DNN is a function $ \mathcal{N}:\mathbb{R}^m\rightarrow\mathbb{R}^k $ which is a composition
of $ n $ layers $ L_0,...,L_{n-1} $. Each layer $ L_i $ consists of
$ s_i \in \mathbb{N} $ nodes, denoted $ v^1_i,...,v^{s_i}_i $. 

The assignment for the
$ j^{th} $ node in the $ 1 \leq i < n-1 $ layer is computed as

\begin{equation}    
 v^j_i =f\left( \underset{l=1} { \overset {s_{i-1}} { \sum } } 
  w_{i,j,l} \cdot v^l_{i-1} + b^j_i \right)
\end{equation}

for some non-linear function $f:\mathbb{R}\rightarrow\mathbb{R}$, called the \emph{activation function}.

The neurons in the last, output layer are computed in a similar manner, without using $f$.
The parameters 
$ w_{i,j,l}$ and $ b^j_i $ are predetermined and are called the weights and biases of $\mathcal{N}$, respectively.
One of the most common activation functions is the
\textit{rectified linear unit} (\relu{}), defined as  $\relu(x) = \max(x,0) $.

\section{Checking theory lemmas.}
\label{app:lemmachecking}
Empirically, the majority of the proof size and the proof checking process is used for storing the theory lemmas and checking them. Thus, we decided to enable two checking modes, allowing different balances of scalability and reliability. The modes are 
\begin{inparaenum}[(i)]
\item a complete checking mode in which lemmas correctness are checked, thus prioritising reliability; and 
\item a partial checking mode in which lemmas claims are used without checking, thus prioritising scalability.
\end{inparaenum}
In both modes, the checker iterates through the list of lemmas and updates $u,l$ locally. 
If the complete checking mode is enabled, the checker needs to check in each iteration, that the Farkas vector $w$ corresponds to the details of the bound invoked the lemma, and that it can be used to by a theory lemma to update the learned bound.
To do so, the checker creates the row vector $w^\intercal \cdot A$, which is equivalent some linear combination $w^\intercal \cdot A \cdot x := 0=\underset{j}{\sum}c_j\cdot x_j$. Suppose the lemma claims to prove some (say, upper) bound of variable $x_i$ to be of value z. Then, as shown in~\cite{IsBaZhKa22} the checker checks that for the equation $x_i = \underset{j\neq i}{\sum}c_j\cdot x_j + (c_i+1)\cdot x_i$, the upper bound of $x_i$ is indeed z. If so, the checker continues to pattern match the lemma to any of the theory lemmas for some piecewise-linear constraints. In order to support a lemma, it is required to be hard coded in the checker.
Note that the bound computed using the checking process uses the bound vectors $u,l$ only, whereas the bound value $z$ can be deduced using bound tightening performed by the DNN verifier, which can be much tighter than of $u,l$. 

The theory lemmas currently supported by our proof checker are all can an be learned using a \relu{} constraint of the form $ f = ReLU(b) = \max(0,b)$.
In some cases, these lemma can also be derived using a linear combination, if the corresponding equation has been introduced (i.e., the equation $f = b$).

The lemmas are: 

\begin{center}
	\begin{minipage}{0.8\linewidth}
		\begin{inparaenum}[(i)]
		 	\item For a positive $l(f)$, $l(b) := l(f)$.\\
		 	\item For a positive $l(b)$, $l(f) := l(b)$.\\
		 	\item For any $u(f)$, $u(b) := u(f)$.\\
		 	\item For non-positive $u(b)$, $u(f) := 0$. \\ 
		 	\item For a positive $u(b)$, $u(f) := u(b)$.\\
		 \end{inparaenum}
	\end{minipage}
\end{center}

\section{Native List v. Sparse Vector Implementation}
\label{app:imandra}

This appendix details the pros and cons of the two data structures use for implementing vectors: native lists and sparse vectors (with underlying map/BST).

\emph{Performance}. Access to random elements, an operation that is often used in the proof checker, is faster for sparse vectors: the complexity for accessing a random element $n$ in a BST is $O(h)$ (where $h$ is the length of the longest path from the root) agaisnt $O(n)$ for the inductive native list definition. This performance advantage can be seen in the benchmark in Table~\ref{tab:results}, Appendix~\ref{app:evaluation}.

\emph{Verification}. Native lists have the benefits of having verified functions and properties available out of the box. However, operations on native lists often have to consider the case of lists of different length. This can be done in the code with error handling, or in the verification phase by adding preconditions on list lengths to all verification goals. As a result code complexity and reasoning difficulty are increased. Sparse lists are total functions, so there is no need to verify dimensions of vectors and matrices; however, we need to prove basic properties before we can start reasoning about them. For now our we have focused our verification effort on native lists.

Listings~\ref{lst:up_nat} and \ref{lst:up_sparse} show the implementation of the same function using native lists and sparse vectors. Notice that the native list implementation uses explicit induction (in the function \lstinline{comput_row_upper_bound'}) and has to include the case where lists have different lenghts in its pattern matching, whereas the sparse vector implementation has more succinct (and more efficient) random element access using \lstinline{M.get}, and the size of vectors is irrelevant.

\begin{figure}
    \centering
\begin{lstlisting}[
                   frame=single, 
                   linewidth=1\textwidth,
                   caption=\footnotesize{\emph{Definition of the function to compute the upper bound of a tableau row using native lists}}, 
                    label=lst:up_nat]
(* compute the upper bound for a tableau row represented as list of reals *)
let compute_row_upper_bound (row: real list) (upper_bounds: real list) (lower_bounds: real list) =
  let rec compute_row_upper_bound' inner_row ub lb res =
      match inner_row, ub, lb with
      | [], [], [] -> res
      | r::rs, u::us, l::ls -> if r <. 0.
        then
          compute_row_upper_bound' rs us ls (res +. (l *. r))
        else
          compute_row_upper_bound' rs us ls (res +. (u *. r))
        (* Handle error cases (list length mismatch) *)
        | _::_, [], [] | [], _::_, [] | [], [], _::_ | [], _::_, _::_ | _::_, [], _::_ | _::_, _::_, [] -> res
  in
  compute_row_upper_bound' row upper_bounds lower_bounds 0.;;
\end{lstlisting}
\end{figure}

\begin{figure}
\begin{lstlisting}[frame=single, 
                   caption=\footnotesize{\emph{Definition of the function to compute the upper bound of a tableau row using sparse vectors}}, 
                    label=lst:up_sparse]
    (* compute the upper bound for a tableau row represented as a M.t / sparse vector *)
let compute_row_upper_bound (row: ('a, 'b) M.t) upper_bounds lower_bounds =
    let linear_comb_keys = row.keys in 
    let acc_func acc var =
        let value = M.get var row in 
        if value <. 0.
        then
            acc +. (M.get var lower_bounds) *. value
        else
            acc +. (M.get var upper_bounds) *. value
    in
    let sum = List.fold_left acc_func 0. linear_comb_keys in
    sum

\end{lstlisting}
\end{figure}

\section{Evaluation on ACAS-Xu Proofs}
\label{app:evaluation}

This table shows the performance for checking proofs generated by Marabou on several verification tasks from the ACAS-Xu benchmark~\cite{KaBaDiJuKo21}. Each task is identified by a network identifier (e.g. N(2,9) and a property number (e.g. p3). The performance for the existing proof checker for Marabou proofs written in C++ \cite{IsBaZhKa22} is compared with our implementation in Imandra. Our implementation is evaluated with different vector implementations (native lists and sparse vectors, as discussed in Section~\ref{subsec:impl}) and in both verification modes, with and without checking the theory lemmas' correctness (as discussed in Section~\ref{sec:construction}). 

\begin{table}[]
    \centering
    \begin{tabular}{P{1.5cm}P{2cm}P{2cm}P{2cm}P{2cm}P{2cm}}
        \toprule
         \textbf{ACAS-} & \textbf{C++\cite{IsBaZhKa22}} & \multicolumn{2}{m{4cm}}{\textbf{Imandra (native lists)}} & \multicolumn{2}{m{4.3cm}}{\textbf{Imandra (sparse vectors)}} \\
         \cmidrule(rl){2-2}\cmidrule(rl){3-4}\cmidrule(rl){5-6}
         \textbf{Xu tasks} & Full & No lemmas & Full & No lemmas & Fulll \\ 
        \midrule
        
        N(2, 9) p3 & 5.130  & 167.078   & 878.075   & 15.125    & 4784.866  \\
        N(2, 9) p4 & 5.658  & 206.675   & 1019.770  & 11.208    & 8817.575  \\
        N(3, 7) p3 & 10.557 & 299.608   & 1493.763  & 24.979    & 1638.844  \\
        N(5, 7) p3 & 2.568  & 58.288    & 311.096   & 50.365    & 12276.323 \\
        N(5, 9) p3 & 15.116 & 424.816   & 2210.472  & 30.611    & 6265.039  \\
    \bottomrule            
    \end{tabular}
    \caption{Comparison of the execution speed for checking a DNN verifier's proofs for verification tasks from the ACAS Xu benchmark.}
    \label{tab:results}
\end{table}

The best mode of our implementation is using sparse vectors with no theory lemmas checking; it is about twice as slow as the original implementation. The best performing full checking of the proof, using native lists, is about 150 times slower than the C++ checker. This performance loss doesn't come as a surprise as the arbitrary precision reals is computationally harder than dealing with fixed precision floating point numbers.

One useful insight for our ongoing optimization work is that the sparse vector mode is faster than the native list mode when theory lemmas are checked, but slower when they are not. This guides us to find inefficient code in the latter configuration. 

\end{document}